\documentclass[aps,prb,onecolumn]{revtex4}

\usepackage{graphicx}
\usepackage{amsfonts}
\usepackage{amsmath,latexsym,amsthm}
\usepackage{float}

\begin{document}

\title{Life in a random universe: Sciama’s argument reconsidered}

\author{Zhi-Wei Wang${}^{1,2\dagger}$
and Samuel L.\ Braunstein${}^{3\ast}$}
\affiliation{${}^1$College of Physics, Jilin University,
Changchun, 130012, People's Republic of China}
\affiliation{${}^2$Department of Physics and Astronomy,
University of Lethbridge, Lethbridge AB  T1K 3M4, Canada}
\affiliation{${}^3$Computer Science, University of York, York YO10 5GH,
United Kingdom }
\affiliation{${}^\dagger$zhiweiwang.phy@gmail.com}
\affiliation{${}^\ast$sam.braunstein@york.ac.uk}

\begin{abstract}

\noindent 
Random sampling in high dimensions has successfully been
applied to phenomena as diverse as nuclear resonances, neural networks
and black hole evaporation. Here we revisit an elegant argument by the
British physicist Dennis Sciama, which demonstrated that were our
universe random, it would almost certainly have a negligible chance for
life. Under plausible assumptions, we show that a random universe can
masquerade as `intelligently designed,' with the fundamental constants
instead appearing to be fined tuned to be achieve the highest
probability for life to occur. For our universe, this mechanism may only
require there to be around a dozen currently unknown fundamental
constants. We speculate on broader applications for the mechanism we
uncover.

\end{abstract}

\maketitle

\section{Introduction} \label{sec:intro}

Whatever might be eventually concluded about a universal definition for
life, we can certainly agree that the universe we inhabit has so far
supported the emergence, evolution and continued sustenance of human
beings. Despite our having grown collectively more powerful than most
known species, within the universe we are very fragile and maintain a
precarious hold on existence. We are carbon based, requiring a planetary
body to live on, which follows a comfortable and steady orbit around a
single and not too energetic star.

These constraints already place tight bounds on the fundamental
constants of the universe. To ensure that a population of yellow dwarf
stars like our sun exist, the fine structure constant must be tuned to
within a percent or two of its current value (outside this narrow range
almost all stars would either be blue giants, or red
dwarfs) \cite{Misner1970}. The cosmological constant must be between
120-124 orders-of-magnitude smaller than its naive quantum
field-theoretic value. The upper bound ensures that bodies can form
gravitationally \cite{Weinberg1987,Weinberg1998} and the lower bound
ensures that nascent life will not be extinguished by proximity to
gamma ray bursts \cite{Piran2016}.
To ensure that nuclear reactions
within stars can form carbon, but not have the process bypassed leaving
only oxygen, a remarkable set of coincidences is required among the
fundamental constants so that one resonant energy level exists, yet
another level just fails to be resonant \cite{Hoyle1954}. Many of the
fundamental constants therefore seem to be boxed into a narrow range of
values compatible with our existence \cite{Carter1974}. Of course,
a less anthropocentric view would considerably broaden this
range \cite{Adams2019} (see discussion).

How can we understand our being in such a human-compatible universe? It
has been suggested \cite{Dicke1961} that the fundamental constants may
have been selected `randomly' among all possible values. If that were
the case, then such compatibility is merely a condition consistent with
our being here to observe it. Given enough potential universes to
randomly choose from, almost anything {\it could\/} happen and the
conditional probability for human compatibility would be one.
Equivalently, if these `random selections' were individual universes
within a multiverse, then our universe being human compatible would be
the same as us being located in one of the universes within the
multiverse where humans are possible. In other words, given enough
potential universes almost anything {\it will\/} happen. Such
`explanations' are said to invoke the weak anthropic
principle \cite{Carter1974}, yet they explain nothing and fail to provide
any real resolution. Are they at least predictive?

Dennis Sciama, considered to be one of the fathers of modern cosmology,
argued that were our universe random, in either sense given above, then
it would almost certainly have a low probability for life as we know
it \cite{Deutsch2011,Wang2023}.

Sciama assumed that the feature distinguishing different potential
universes was the set of specific values taken by the fundamental
constants; the underlying physical laws themselves being fixed. We can
then envision the human-compatible universes as an `island' within a
`sea' of more general possibilities. Each point on the island or in the
sea describes a unique universe that is described by a distinct set of
fundamental constants. The dimensionality of this space of points is
naively given by the number of fundamental constants. Thus the
human-compatible island of universes corresponds to some shape in a high
dimensional space. The shoreline of the island corresponds to the
boundary separating universes with a chance for human life to form from
those where this is impossible. Thus, the shoreline itself will be made
up of universes with an exactly vanishing probability for such life.
Assuming continuity, as one moves inland, this probability will increase,
reaching a maximum presumably somewhere far in from the shoreline.

This probability landscape is different from the chance of randomly
selecting a universe. Because the range of parameters consistent with
human life is quite small, one might expect any smooth measure for
randomly selecting universes, to be approximately uniform across the
island. Sciama's argument now follows from a well-known
concentration-of-measure phenomenon \cite{Milman1988}:

Consider an $n$-dimensional cube (a hypercube) with side length $s$ and
hence volume $s\times s\times \cdots \times s = s^n$. Now suppose you
paint this hypercube, causing the side length to marginally increase to
$s+\delta$. The volume of the paint used is simply $(s+\delta)^n -s^n
\ge n\, s^{n-1}\delta$. But even if the layer of paint is very thin, so
$\delta \ll s$, in sufficiently high dimensions, $n\ge s/\delta$, the
volume of the paint will exceed the volume of the original hypercube.
Thus, were we to grind up our thinly painted hypercube and take a random
sample, we would most likely find paint!

\begin{figure}[ht]
\centering
\vskip -0.1truein
\includegraphics[width=0.7\textwidth]{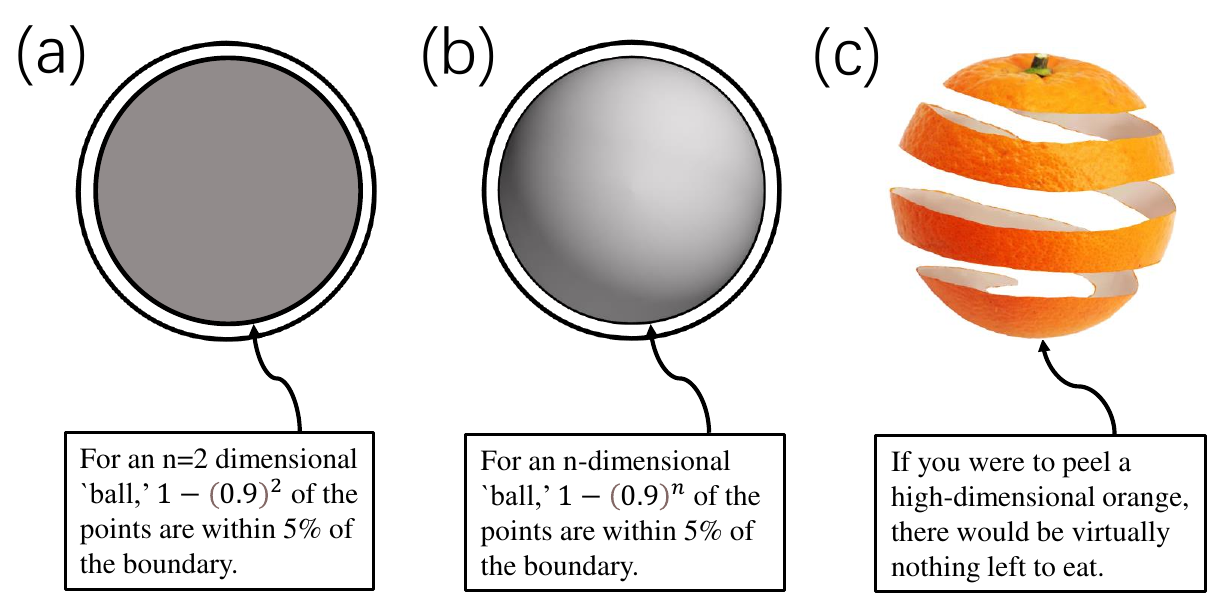}
\vskip -0.1truein

\caption{An `unappealing' consequence of concentration-of-measure
phenomena: (a) an $n=2$ dimensional `ball' has $1-(0.9)^2=0.19$ fraction
of the points within $5\%$ of the boundary; (b) an $n=3$ dimensional
ball has a fraction $1-(0.9)^3=0.271$ within $5\%$ of the boundary; for
large $n$, the fraction $1-(0.9)^n$ approaches unity; therefore, (c) if
you were to peel a high-dimensional orange, there would be almost
nothing left behind!}

\vskip -0.1truein
\label{fig0}
\end{figure}

This is not only true for hypercubes, but for any shape in high
dimensions \cite{Milman1988}. The volume, and similarly the weight, will
be entirely concentrated within a thin layer at the surface. Thus,
figuratively, a high-dimensional orange is essentially only its peel.
See Fig.~\ref{fig0}.

Applied to the high-dimensional island of human-compatible universes, a
randomly selected universe will then almost certainly be found in a
narrow band on the shore. However, since there is no chance for life at
the shoreline or anywhere off the island, the probability for life would
be expected to be very low for a typical random universe lying in this
narrow band at the shore.

This prediction is in contrast to that of intelligent design where one
might expect a universe further inland closer to, or possibly achieving,
the greatest chance for human life.

\section{Method} \label{sec:Method}

Is this space of universes really high dimensional? In 1936 Eddington
counted four fundamental constants \cite{Eddington1936}. This count
excludes Newton's gravitational constant, the speed of light, Planck's
constant and the permittivity of the vacuum, all used to provide scales
for dimensional quantities like length, time, mass and electric
charge \cite{Eddington1936}. Just a few years ago this count had grown to
26 for the `standard model' including the cosmological constant for
gravity \cite{Siegel2018}. Today, if we add three neutrino masses, the
count would be 29. However, our current model of the universe hardly
explains everything. There remain numerous long-standing open questions,
many cosmological in nature, such as matter-antimatter asymmetry, dark
matter, dark energy and more. Thus it would be surprising if the total
number of the fundamental constants in a complete theory of the universe
were not much larger.

Although it did not figure into Sciama's original argument, we shall see
that the shape of the island plays a crucial role in the possible
apparent reversal of Sciama's conclusion. Note that the `orange peel'
result itself is essentially independent of this shape, which follows
simply from the scaling of the `hypervolume' with dimensionality. Thus,
there is no question about a randomly selected universe compatible with
human life having a set of fundamental constants that almost certainly
lie on the narrow shore, with a low chance for life like us.

Notwithstanding this, where on the island the universe appears to lie
can depend on the island's shape. This can be the case whenever our
knowledge of the list of fundamental constants is incomplete. In this
case, we would consider the island and its surrounding sea to be a
lower-dimensional space than it actually is. Our view of the island
would be one that projects out the unknown constants. This may be
visualized as an `X-ray' of the actual $n$-dimensional island onto a
lower $m$-dimensional island describing the known constants. See
Fig.~\ref{fig1}.

\section{Results} \label{sec:Results}

\begin{figure}[ht]
\centering
\vskip -0.1truein
\includegraphics[width=0.8\textwidth]{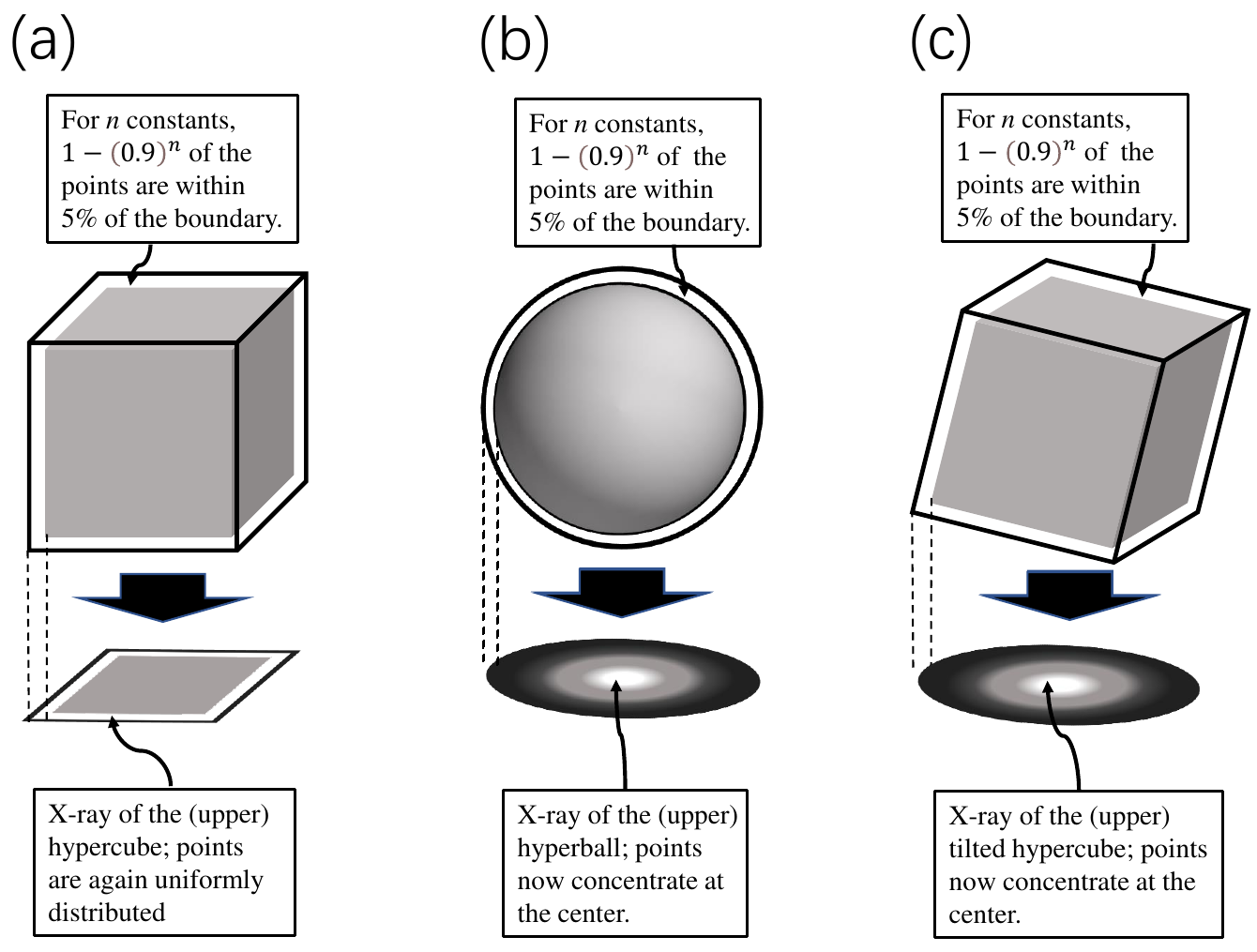}
\vskip -0.1truein

\caption{Random sampling from the human-compatible island can look
different dependent on its shape, if one has only limited access to the
fundamental constants. The high-weight region (typically the `shore') is
shown in white, with the remaining low-weight contribution in gray. The
island of the accessible fundamental constants is obtained by
integrating out those constants which are unknown or unobserved; this is
visualized as an `X-ray' of the actual island. (a) For an island which
is an $n$-dimensional hypercube (upper), an X-ray reduces to a
uniform-measure hypercube in a lower dimension (lower). (b) For an
island which is a uniformly-distributed $n$-dimensional ball (upper)
with many unknown constants, its X-ray is well approximated by a narrow
Gaussian concentrated at the center of the human-compatible island
(lower). (c) For an island which is an $n$-dimensional
hypercube (upper), an X-ray along a randomly oriented direction is again
well approximated by a Gaussian concentrated at the center of the
human-compatible island (lower).}

\vskip -0.1truein
\label{fig1}
\end{figure}

In the case that the island has the shape of a uniform-weight
$n$-dimensional cube (a hypercube), with independent bounds on each
constant, the X-ray is simply a lower-dimensional uniform-weight
hypercube. See Fig.~\ref{fig1}(a). Again the lower-dimensional shore
contains the greatest weight.

By contrast, for a uniform-weight hyperball shaped island,
the X-ray, integrating out many dimensions, leads to a narrow Gaussian
with the weight concentrated at the center of the island, far inland
from the shoreline (see Appendix). See
Fig.~\ref{fig1}(b). Further, if the uniform-weight
$n$-dimensional hypercube is X-rayed along a skewed orientation (e.g.,
one randomly chosen) the projected island is again well approximated by
a Gaussian with the weight concentrated at the center of the island (see
Appendix). See Fig.~\ref{fig1}(c).

Surprisingly, the result we find for a hyperball-shaped island shown in
Fig.~\ref{fig1}(b), or equivalently, the skew-oriented
hypercube-shaped island shown in Fig.~\ref{fig1}(c) may well
represent the generic result. 

Firstly, the human-compatible island will be formed by
those universes whose fundamental constants simultaneously satisfy a
series of human-compatible constraints. Each of these constraints may be
thought of as dividing the space of all possible universes into two
subsets: those that satisfy a specific constraint for life and those
that fail to. Now, as already mentioned, the range of parameters
consistent with human life is quite small; the island itself is in some
sense `small.' Consequently, assuming each constraint is smooth, its
action constraining our island should be well approximated locally by a
separating hyperplane in the space of universes. Combining the
hyperplanes of these individual human-compatible constraints then yields
a description of the human-compatible universes as well approximated by
a convex faceted island.

Secondly, the various correlations and coincidences found
among the fundamental constants when determining the human-compatible
island's shoreline \cite{Hoyle1954,Carter1974,Eddington1939},
suggest that the facets associated with such correlations
will be tilted with respect to the axes given by the fundamental
constants themselves. Combined, these arguments yield a convex island
whose facets have a skewed orientation.

Finally, the projective central limit
theorem \cite{Diaconis1984,Klartag2007} ensures that
virtually any projection of such a high $n$-dimensional uniform-weight
shape will be well approximated by a Gaussian with variance scaling as
$1/n$ (with respect to a suitably chosen diameter).

\begin{figure}[ht]
\centering
\vskip -0.1truein
\includegraphics[width=0.5\textwidth]{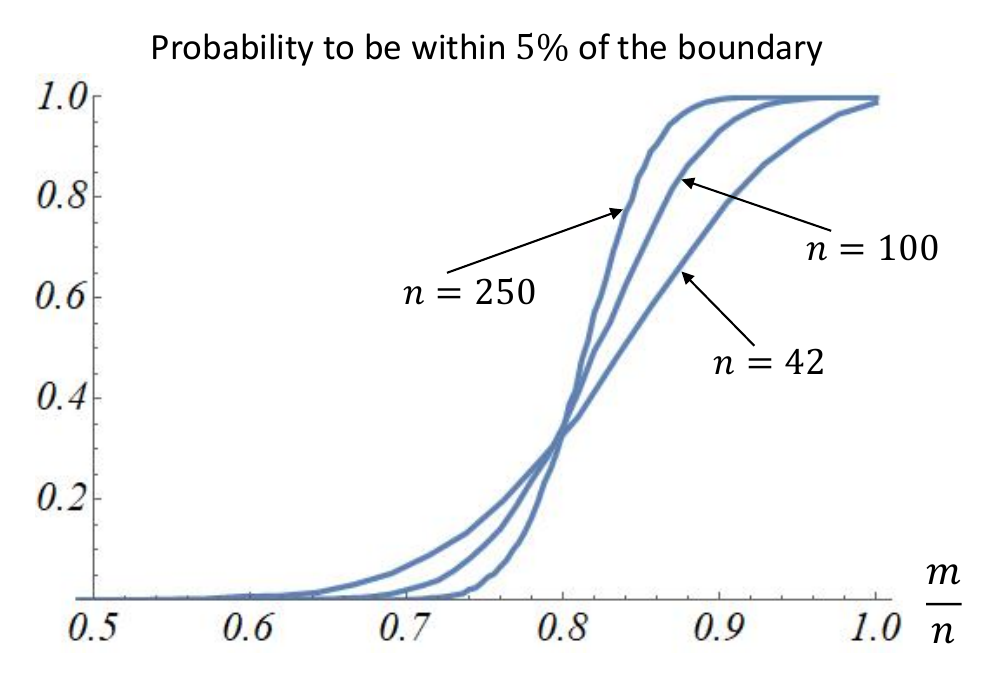}
\vskip -0.1truein

\caption{Probability for the $m$ accessible constants of a random
universe to be within $5\%$ of the life-denying boundary, versus the
accessible fraction $m/n$. The calculation assumes the island has the
shape of a hyperball with $n$ fundamental constants, though the
existence of only $m$ is known. We consider $n\in\{42,100,250\}$. In
each case, unless at least $80\%$ of the total number of constants are
accessible ($m/n\ge0.8$), the chance of being near the boundary is less
than $\simeq 0.35$.}

\vskip -0.1truein
\label{fig2}
\end{figure}

However, the projective central limit theorem only tells us that the
distribution is peaked far inland from the accessible (projected)
shoreline as $n\rightarrow\infty$ (for the hyperball, in fact we find
that $m/n \ll 1$ is sufficient; see Appendix). What
about the probability for the accessible parameters nevertheless being
found on the projected shore near the boundary? We find a universal
behavior for the tail of the distribution of the projected hyperball as
$m/n \rightarrow 1$. In particular, in Fig.~\ref{fig2} we compute the
probability to be within $5\%$ of the projected shoreline versus the
fraction of accessible constants, $m/n$, when an $n$-dimensional
hyperball is projected down to $m$ dimensions. For $n\in \{42, 100,
250\}$, we see that if less than a threshold of around
$80\%$ of the total number of fundamental constants are accessible, then
the chance of being near the boundary is less than around $0.35$. Thus,
taking as null hypothesis that our universe is random, there would be a
low chance for finding the fundamental constants of the universe to be
near the shoreline until we had knowledge of the vast majority of all
the universe's parameters.

\vskip 0.1in

\section{Discussion and Summary} \label{sec:Discussion}

In summary, Sciama's reasoning suggests that were our universe random,
there would be a statistical signature whereby the fundamental constants
would almost certainly lie near the boundary of human-compatible
universes.

One might view Sciama's result to be solely that a random universe would
lead to a scenario where life as we know it is only barely possible.
This `orange peel' argument stands firm and may even explain the
apparent scarcity of intelligent life in the universe, potentially
resolving Fermi's paradox \cite{Cirkovic2016}. However, since the island
of parameters consistent with any type of lifeform would appear to be
significantly larger \cite{Adams2019} than that considered solely for the
sake of humans, it is likely that life itself may be very common in our
universe. This rough-and-ready prediction for a random universe may well
be falsifiable within the coming years.

However, presuming that our knowledge of the fundamental constants is
incomplete, we have shown the signature for a random universe can be
reversed. For example, were our universe random with 42 fundamental
constants (instead of the merely 29 currently known), and taking the
shape of the human-compatible island of parameters as discussed above,
there would be only $\simeq5.5\%$ chance of the set of those 29
currently known constants to lie within $5\%$ of the boundary where
human life becomes impossible. Instead, the greatest likelihood would be
to find these known constants to be far within the `projected'
human-compatible island of universes, mimicking a universe built by
intelligent design to create intelligence.

Currently there is no direct evidence to support the claim that Sciama's
statistical signature applies to our universe (outside of consistency
with the null results from SETI \cite{SETI}). However, this observation
is in the context of fundamental theories which cannot yet explain
everything about our universe, so there is a widely accepted expectation
that new physics, along with additional fundamental constants, would be
needed. Further, our current best guess for a fundamental theory, string
theory, naturally contains a multiverse and hence a random selection
mechanism. Combined with our analysis above, these reasoned expectations
suggest the statistical prediction that \emph{at least} around a dozen
fundamental constants, and possibly many more, are yet to be discovered
to fully explain our universe.

Can any wider scientific lesson be learned from our
arguments? Beyond the anthropic issues discussed here, they may be
relevant to astronomy and indeed any field when viewed as a data
science. After all, our analysis suggests potential pitfalls when
considering how to interpret data sampled from constrained sets. In
particular, unknown degrees of freedom are common in some systems and
the viewing of low-dimensional `sections' of high-dimensional data is
virtully the {\it sine qua non\/} of any data science strategy.

Finally, humanity's looking out to the stars has always had
at its heart a search for meaning. Curiously, it may well be that our
analysis here shows, {\it not\/} how to create meaning `out of whole cloth,'
but how to greatly enhance any scintilla.

Indeed, if we consider the intelligent design of the
universe as an artful act, whatever else it might be, then we have
uncovered a mechanism whereby even a random universe may appear artful;
or loosely speaking, whereby even an atheist might say \cite{Wilde1889} ``life imitates
art.'' Recalling that our analysis is based on
concentration-of-measure phenomena in high-dimensional spaces, it is
natural to ask whether this mechanism for imitating art may not have
grander application. For example, could one enhance the
artfulness of an almost soullessly generated piece of `art' to make it
mimic a true work of art; not by slavish copying, but perhaps by so
constraining the work around with interconnections and correlations --
the coincidences constraining our fundamental constants -- that one may
begin to find it harder to distinguish between such a piece and an
intelligently crafted work? These correlations acting to
`tilt the facets of our island' and so produce an apparent
lower-dimensional artful enhancement. For something complex, having a
sufficient number of working parts, it may not even be necessary to hide
part of the work; or as with our analysis here, it may be crucial to
first build the work and then make large portions inaccessible; be they
backstory, foundation, milieu, history, whatever. And if anything like
this can succeed, why not look further, perhaps toward the imitation of
intelligence itself. Maybe the magic behind creating meaning has
`simply' been a matter of hiding much of the supporting artifice from
the audience, and even from ourselves.

\section*{Acknowledgments}

Z.-W.W. wishes to express his sincere gratitude to Professors Yanming Ma and Haijun Wang for their gracious invitation and invaluable support throughout this project.

\section*{APPENDIX}

\subsection{Projection of an ${(n-1)}$-sphere to one dimension}

Consider a smooth convex $n$-dimensional geometric body with a uniform
probability distribution across and within it. The projective central
limit theorem claims that as $n\to\infty$, if we project such a body to
lower dimensions, the probability will concentrate to a `center' of
the lower dimensional object \cite{Diaconis1984,Klartag2007,Knaeble2015Ap}.
This phenomenon has been proved for all smooth convex geometric bodies
and the limiting probability distribution is claimed to be a Gaussian
distribution with variance scaling as $1/n$.

Here we compute this exactly for an $(n-1)$-sphere (the surface of an
$n$-dimensional ball) projected to $m$ dimensions. We find for large
$n-m$ that the resulting distribution is Gaussian with variance scaling
as $1/n$. This is in agreement with the more general, though
looser claim of the projective central limit theorem when $n\gg m$.
Finally, combining the `orange peel' concentration of measure
result, we argue that the same limit to a Gaussian with variance scaling as
$1/n$ will hold for the projection of an $n$-dimensional ball onto
$m$ dimensions.

Because the general calculation is rather complicated, we start with
the simpler case of projecting an $(n-1)$-sphere onto a single dimension.

An $(n-1)$-sphere with unit radius in n-dimensional Euclidean space
(Cartesian coordinates) may be described as satisfying $x^2_1+x^2_2+\cdots
+x^2_n=1$. It may be transformed into the hyper-spherical coordinate by
\begin{eqnarray}
x_1 &=& -\cos \varphi_1 \nonumber \\
x_2 &=& -\sin \varphi_1 \cos \varphi_2 \nonumber \\
&\vdots & \nonumber \\
x_{n-1} &=& -\sin \varphi_1 \sin \varphi_2 \cdots \cos \varphi_{n-1}
\nonumber \\
x_n &=& -\sin \varphi_1 \sin \varphi_2 \cdots \sin \varphi_{n-1} ,
\label{coordinate}
\end{eqnarray}
where $\varphi_1, \varphi_2, \ldots, \varphi_{n-2} \in [ 0,\pi ]$ and
$\varphi_{n-1} \in [0,2\pi]$. Here the minus sign is just for future
convenience; note that each $x_i$ is not sensitive to such a minus sign.

Then in hyper-spherical coordinates, the volume element of such an
$(n-1)$-sphere may be written
\begin{equation}
d\Omega_{n-1} = \sin^{n-2} (\varphi_1) \sin^{n-3} (\varphi_2) \cdots
 \sin^2 (\varphi_{n-3})  \sin (\varphi_{n-2}) \, d\varphi_1 d\varphi_2
 \cdots d \varphi_{n-2} d\varphi_{n-1} ,
\label{nsphere}
\end{equation}
as is easily checked by computing the Jacobian for this transformation.
Integrating this volume element over the entire sphere yields the
standard result
\begin{eqnarray}
S_{n-1} &=&\int d\Omega_{n-1} \nonumber \\
    &=& \int\sin^{n-2} (\varphi_1) \sin^{n-3} (\varphi_2)
 \cdots  \sin^2 (\varphi_{n-3})  \sin (\varphi_{n-2})
 \, d\varphi_1 d\varphi_2 \cdots d \varphi_{n-2} d\varphi_{n-1} \nonumber \\ 
    &=& \pi^{\frac{n-2}{2}} \frac{\Gamma(\frac{n-1}{2})}{\Gamma(\frac{n}{2})}
\frac{\Gamma(\frac{n-2}{2})}{\Gamma(\frac{n-1}{2})}
  \cdots  \frac{\Gamma(\frac{2}{2})}{\Gamma(\frac{3}{2})} 2\pi \nonumber \\ 
    &=& \frac{2 \pi^{\frac{n}{2}} }{\Gamma(\frac{n}{2})}  ,
\end{eqnarray}
where $\Gamma(n)$ is the gamma function so $\Gamma(1)=1$, and in
moving from the second to the third line we have used the result
that
\begin{equation}
\int_0^\pi
\sin^m(\varphi)\, d\varphi = \sqrt{\pi}\,
\frac{\Gamma(\frac{m+1}{2})}{\Gamma(\frac{m+2}{2})}.
\end{equation}

We now consider projecting the uniformly distributed $(n-1)$-sphere
onto a single dimension (i.e., the case $m=1$).
Normalizing the measure of Eq.~(\ref{nsphere}) by $S_{n-1}$, we may compute
the expectation of a general function of $\varphi_1$ as
\begin{eqnarray}
\langle f(\varphi_1)\rangle
&=&\frac{1}{S_{n-1}}\int f(\varphi_1) d\Omega_{n-1} \nonumber \\
    &=& \frac{1}{S_{n-1}} \int^\pi_0 f(\varphi_1) \sin^{n-2} (\varphi_1)
 \, \pi^{\frac{n-3}{2}}  \frac{\Gamma(\frac{n-2}{2})}{\Gamma(\frac{n-1}{2})}
  \frac{\Gamma(\frac{n-3}{2})}{\Gamma(\frac{n-2}{2})}
\cdots  \frac{\Gamma(\frac{2}{2})}{\Gamma(\frac{3}{2})}
2\pi \, d\varphi_1   \nonumber \\
    &=& \frac{\Gamma(\frac{n}{2}) }{2 \pi^{\frac{n}{2}}}
\int^\pi_0 f(\varphi_1) \sin^{n-2} (\varphi_1) \, 
 \frac{2 \pi^{\frac{n-1}{2}}}{\Gamma(\frac{n-1}{2})} \, d\varphi_1 
 \nonumber \\ 
    &=& \int^\pi_0 f(\varphi_1) \sin^{n-2} (\varphi_1) \, 
 \frac{ \Gamma(\frac{n}{2}) }{\sqrt{\pi}\, \Gamma(\frac{n-1}{2})}
\, d\varphi_1  .
\label{shpereP1}
\end{eqnarray}
Consequently, the distribution on $\varphi_1$ is given by
\begin{equation}
\text{P} (\varphi_1)\, d\varphi_1=
\sin^{n-2} (\varphi_1) \, 
 \frac{ \Gamma(\frac{n}{2}) }{\sqrt{\pi}\, \Gamma(\frac{n-1}{2})}
\, d\varphi_1 , \qquad \varphi_1\in [0,\pi]
\label{distP1}
\end{equation}
To see the probability distribution in Euclidean space, we need to
transform Eq.~(\ref{distP1}) back to the coordinate
$x_1$. Since $x_1=-\cos\varphi_1$, we have $dx_1=\sin \varphi_1
d\varphi_1$ and $\sin(\varphi_1)=\sqrt{1-x^2_1}$ and hence we obtain
\begin{equation}
\text{P} (x_1)\, dx_1=
\frac{ \Gamma(\frac{n}{2}) }{\sqrt{\pi}\,
\Gamma(\frac{n-1}{2})} (1-x_1^2)^{\frac{n-3}{2}} dx_1, \qquad x_1 \in [-1,1].
\label{shpereP3}
\end{equation}
We may also obtain an exact expression for the variance
\begin{equation}
(\Delta x_1)^2 = \frac{1}{n}.
\end{equation}

It is easy to see that this probability distribution gets narrower and
narrower as $n$ increases. Using the result that $\lim_{n \to \infty}
(1-\frac{x}{n})^n=e^{-x}$ we see that for sufficiently small $x_1$ and
large $n$ Eq.~(\ref{shpereP3}) may be approximated by a Gaussian with
variance $1/n$, for the case $m=1$. This result is in exact agreement
with that given in previous work \cite{Knaeble2015Ap}, we shall see below
that the exact result yields a subtly different outcome for $m>1$.

\subsection{Projecting to ${m}$-dimension}

When projecting an $(n-1)$-sphere onto $m$ Cartesian dimensions, the
logic is similar to that given in the previous section but now we will
need to integrate out the angles from the set $\{\varphi_{m+1},
\varphi_{m+2}, \ldots, \varphi_{n-1}\}$. Thus, we find
\begin{eqnarray}
\!\!\!&&\langle f(\varphi_1,\ldots,\varphi_m)\rangle
=\frac{1}{S_{n-1}}\int f(\varphi_1,\ldots,\varphi_m)\,d\Omega_{n-1} \nonumber \\
    &=& \frac{1}{S_{n-1}} \int^\pi_0 \cdots  \int^\pi_0
f(\varphi_1,\ldots,\varphi_m)\,
\sin^{n-2} (\varphi_1) \sin^{n-3} (\varphi_2)
\cdots \sin^{n-m-1} (\varphi_m) \nonumber \\
&&\phantom{\frac{1}{S_{n-1}} \int^\pi_0 \cdots  \int^\pi_0}
\times \pi^{\frac{n-m-2}{2}}
 \frac{\Gamma(\frac{n-m-1}{2})}{\Gamma(\frac{n-m}{2})}
 \frac{\Gamma(\frac{n-m-2}{2})}{\Gamma(\frac{n-m-1}{2})}
\cdots  \frac{\Gamma(\frac{2}{2})}{\Gamma(\frac{3}{2})}\,
2\pi \, d\varphi_1 d\varphi_2 \cdots d\varphi_m  \nonumber \\
    &=& \frac{\Gamma(\frac{n}{2}) }{2 \pi^{\frac{n}{2}}}
\int^\pi_0 \cdots  \int^\pi_0 f(\varphi_1,\ldots,\varphi_m)\,
\sin^{n-2} (\varphi_1) \sin^{n-3} (\varphi_2) \cdots \sin^{n-m-1}
(\varphi_m) \, \frac{2 \pi^{\frac{n-m}{2}}}{\Gamma(\frac{n-m}{2})}
\, d\varphi_1 d\varphi_2 \cdots d\varphi_m  \nonumber \\ 
    &=& \int^\pi_0 \cdots  \int^\pi_0 f(\varphi_1,\ldots,\varphi_m)
\sin^{n-2} (\varphi_1) \sin^{n-3} (\varphi_2) \cdots
\sin^{n-m-1} (\varphi_m) \,
  \frac{ \Gamma(\frac{n}{2}) }{\pi^{\frac{m}{2}}
\Gamma(\frac{n-m}{2})} \, d\varphi_1 d\varphi_2 \cdots d\varphi_m  .
\label{mshpereP1}
\end{eqnarray}
Since the volume element described by two different coordinates systems
may be connected by the Jacobian, we have 
\begin{equation}
dx_1 dx_2 \cdots dx_m = \text{J} \, d\varphi_1 d\varphi_2 \cdots d\varphi_m ,
\label{Ctransf1}
\end{equation}
where J is the Jacobian describes this volume transformation.
From Eq.~(\ref{coordinate}), we know that J may be written as
\begin{eqnarray}
\text{J} = \Bigl | \frac{\partial x_i}{\partial \varphi_j} \Bigr |  =
\left| \begin{array}{cccc}
\sin\varphi_1 & 0 & \cdots & 0 \\
\cos\varphi_1\cos\varphi_2 & \sin\varphi_1\sin\varphi_2 & \cdots & 0 \\
\vdots & \vdots & \ddots & \vdots \\
\cos\varphi_1\sin\varphi_2\cdots\sin\varphi_{m-1}\cos\varphi_m & \sin\varphi_1\cos\varphi_2\cdots\sin\varphi_{m-1}\cos\varphi_m & \cdots & \sin\varphi_1\cdots\sin\varphi_m
\end{array} \right|
\label{Jac1}
\end{eqnarray}
Since terms in the upper triangle in the Jacobian, Eq.~(\ref{Jac1}),
are all zero, the Jacobian trivially reduces to
\begin{eqnarray}
\text{J} = \sin ^m (\varphi_1) \sin ^{m-1} (\varphi_2)
\cdots \sin (\varphi_m) .
\label{Jac2}
\end{eqnarray}
Inserting Eq.~(\ref{Jac2}) into Eq.~(\ref{Ctransf1}) yields
\begin{equation}
dx_1 dx_2 \cdots dx_m = \sin ^m (\varphi_1) \sin ^{m-1} (\varphi_2)
\cdots \sin (\varphi_m) \, d\varphi_1 d\varphi_2 \cdots d\varphi_m ,
\label{Ctransf2}
\end{equation}
and substituting Eq.~(\ref{Ctransf2}) into Eq.~(\ref{mshpereP1}) yields
\begin{eqnarray}
&&\langle f(\varphi_1,\ldots,\varphi_m) \rangle \nonumber \\
&=&\int^1_{-1} \int_{-\sin \varphi_1}^{\sin \varphi_1}
\cdots  \int_{-\sin (\varphi_1) \cdots
\sin (\varphi_{m-1})}^{\sin (\varphi_1)
\cdots \sin (\varphi_{m-1})}f(\varphi_1,\ldots,\varphi_m)\,
 \sin^{n-m-2} (\varphi_1)
\sin^{n-m-2} (\varphi_2) \cdots \sin^{n-m-2} (\varphi_m) \nonumber \\
&&\phantom{\int^1_{-1} \int_{-\sin \varphi_1}^{\sin \varphi_1}
\cdots  \int_{-\sin (\varphi_1) \cdots
\sin (\varphi_{m-1})}^{\sin (\varphi_1)
\cdots \sin (\varphi_{m-1})} }
  \times \frac{ \Gamma(\frac{n}{2}) }{\pi^{\frac{m}{2}}
\Gamma(\frac{n-m}{2})} \, dx_1 dx_2 \cdots dx_m  , \nonumber \\
\label{mshpereP2}
\end{eqnarray}
where $\sin (\varphi_i)$ is positive function of $x_i$.

To further simplify Eq.~(\ref{mshpereP2}), let us first consider
$x^2_1 + x^2_2 + \cdots + x^2_m$. From Eq.~(\ref{coordinate}), this
may be written as
\begin{eqnarray}
x^2_1 + x^2_2 + \cdots + x^2_m 
&=& \cos^2 (\varphi_1) + \sin^2 (\varphi_1) \cos^2 (\varphi_2)
+ \cdots + \sin^2 (\varphi_1) \sin^2 (\varphi_2)
\cdots \cos^2 (\varphi_m) \nonumber \\ 
&=& \cos^2 (\varphi_1) + \sin^2 (\varphi_1) ( 1-\sin^2 (\varphi_2) )
+ \cdots + \sin^2 (\varphi_1) \sin^2 (\varphi_2)
\cdots \cos^2 (\varphi_m) \nonumber \\ 
&=& 1 - \sin^2 (\varphi_1) \sin^2 (\varphi_2)  + \cdots +
\sin^2 (\varphi_1) \sin^2 (\varphi_2) \cdots \cos^2 (\varphi_m) .
\end{eqnarray}
The above procedure can be repeated until we arrive at
\begin{eqnarray}
x^2_1 + x^2_2 + \cdots + x^2_m = 1 - \sin^2 (\varphi_1) \sin^2 (\varphi_2)
\cdots \sin^2 (\varphi_m) .
\label{x2x2}
\end{eqnarray}
Applying Eq.~(\ref{x2x2}) to Eq.~(\ref{mshpereP2}) then gives
\begin{eqnarray}
&& \langle f(x_1,\ldots,x_m)\rangle\nonumber \\
&=& \int^1_{-1} \int_{-\sqrt{1-x_1^2}}^{\sqrt{1-x_1^2}} \cdots 
\int_{-\sqrt{1-x_1^2-x_2^2-\cdots -x_{m-1}^2}}
^{\sqrt{1-x_1^2-x_2^2-\cdots -x_{m-1}^2}}
f(x_1,\ldots,x_m)\,
\frac{ \Gamma(\frac{n}{2}) }{\pi^{\frac{m}{2}}
f(x_1,\ldots,x_m)
\Gamma(\frac{n-m}{2})}  \Bigl( 1-\sum_{i=1}^{m} x_i^2
\Bigr)^{\frac{n-m-2}{2}} dx_1 dx_2 \cdots dx_m  , \nonumber \\
\label{mshpereP3}
\end{eqnarray}

By spherical symmetry, it is sufficient to compute the variance on $x_1$, but
this trivially reduces to the result already obtained, since to compute
it we may integrate out all the remaining coordinates $x_2,\ldots,x_m$.
Therefore we find exactly
\begin{eqnarray}
\langle x_i \rangle &=& 0 \nonumber \\
\langle x_i x_j \rangle &=& \delta_{ij} (\Delta x_i)^2 
=\frac{\delta_{ij}}{n}.
\end{eqnarray}
From similar reasoning, for sufficiently large $n$ the distribution 
becomes Gaussian.

Therefore, the probability distribution over this reduced m-sphere reduces
to
\begin{equation}
\text{P} (x_1, x_2, \cdots, x_m) =
\frac{ \Gamma(\frac{n}{2}) }{\pi^{\frac{m}{2}}
\Gamma(\frac{n-m}{2})} \,
\Bigl( 1-\sum_{i=1}^{m} x_i^2 \Bigr)^{\frac{n-m-2}{2}} ,
\end{equation}
with suitable limits on the $x_i$.
This is for the projection of an $(n-1)$-sphere to an $m$-dimensional
subspace.

The limit of this distribution as ${n-m} \to \infty$ and sufficiently
small $x_i$, $i\in\{1,\ldots,m\}$, may be approximated by a Gaussian
with mean zero and a variance in every direction of $1/n$. This result
agrees with previous work \cite{Knaeble2015Ap}, except on the condition
needed, here $n-m$ large as opposed to merely $n$ being large. The
difference in this requirement means that as $m\to n$, where we
project out fewer and fewer coordinates and finally none, the Gaussian
approximation is found to wholly fail.

\subsection{Projection of a rotated hypercube to two dimensions}

\begin{figure}[H]
\centering
\includegraphics[width=0.6\textwidth]{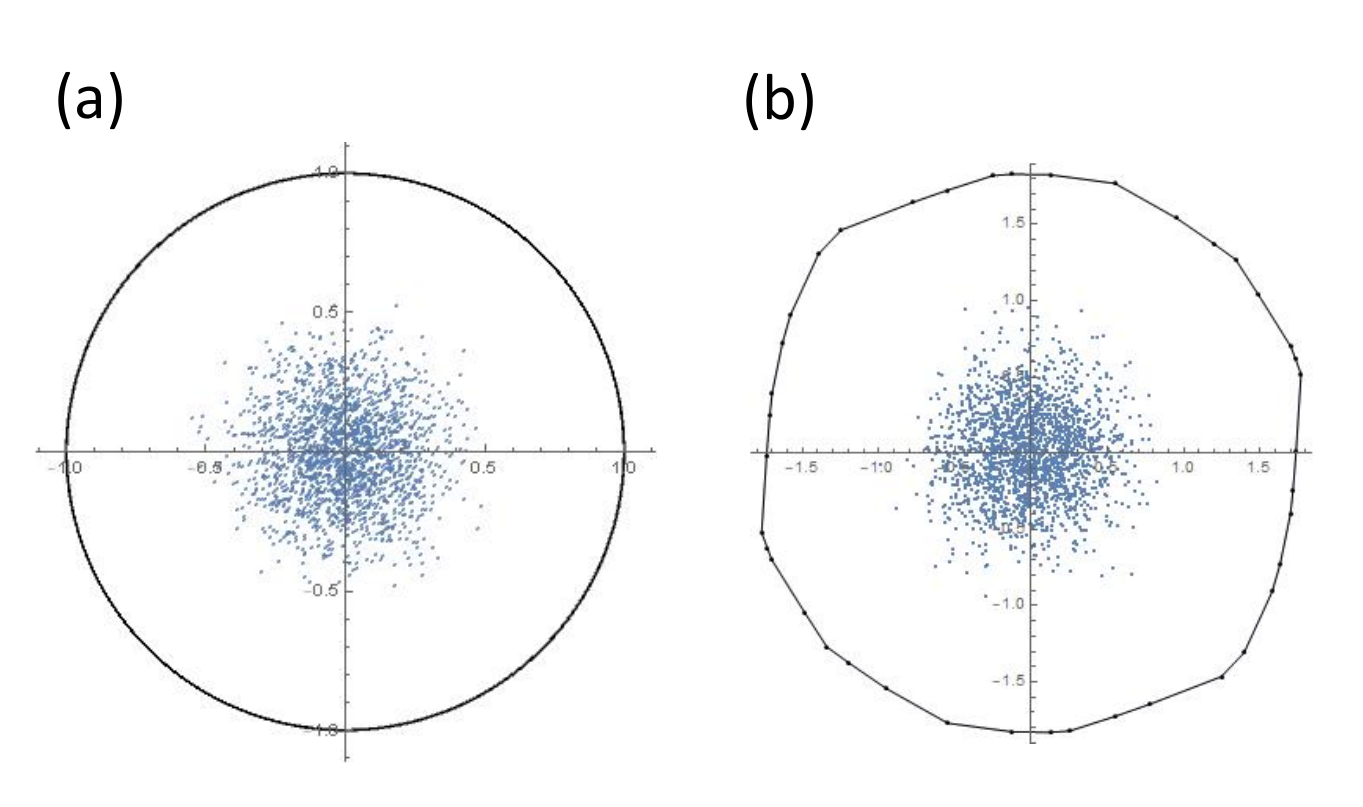}
\vskip -0.1truein

\caption{Plots of both the X-ray (as a scatter plot of 2000
points) where a higher density of points represents a larger weight of
material being X-rayed and the `shadow' outline projection onto two
dimenstions of a (a) 33-dimensional hyperball and a (b) randomly
oriented 20-dimensional hypercube. For the shadow of the hypercube, the
dots along the outline represent the location of the corners there.}

\vskip -0.1truein
\label{fig4}
\end{figure}

Our analysis, both analytic and numeric has largely been based on the
projection of high-dimensional hyperballs. Here we consider the
numerical projection of a randomly rotated high-dimensional hypercube to
two dimensions. Although computing the X-ray of such a hypercube is
straightforward, computing the projection of the boundary very quickly
becomes computationally inaccessible. In particular, one needs to take
the $2^n$ corners of the hypercube and project them to the lower
dimension of interest and then compute the convex hull of those
projected corners. This convex hull represents the `shadow' of the
rotated hypercube under ordinary light (which is assumed to be unable to
penetrate the object itself). In Fig.~\ref{fig4} we illustrate this
computation, comparing both the X-rays and shadows projected onto two
dimensions of (a) a 33-dimensional hyperball and (b) a randomly rotated
20-dimensional hypercube (the highest dimension we could compute in a
reasonable amount of time).

As can be seen, in sufficiently high dimensnions the shadow of a
randomly rotated hypercube, Fig.~\ref{fig4}(b), looks remarkably like
that of a hyperball, Fig.~\ref{fig4}(a). Further, up to defining a
suitable `diameter' to make a more rigorous comparison, the X-ray
projection of such hypercubes appears remarkably similar to that of the
hyperball (again see Fig.~\ref{fig4}). Consequently, in our main
manuscript we used a hyperball to compute the threshold illustrated by
Fig.~3 there in even higher dimensions than we can manage for
hypercubes.

Were much larger computational facilities available a more careful
comparison should be possible, though even only a 42-dimensional
hypercube entails over 4 trillion corners, so computing the projected
boundary into any dimension greater than one \cite{Knaeble2015Ap}
would require major
computational effort.


\begin{thebibliography}{99}

\bibitem{Misner1970}
C.\ W.\ Misner, K.\ S.\ Thorne \& J.\ A.\ Wheeler,
{\it Gravitation},
(Freeman \& Co., San Francisco, 1973) p.~1216.

\bibitem{Weinberg1987}
S.\ Weinberg,
Anthropic bound on the cosmological constant
{\it Physical Review Letters\/} {\bf 59}, 2607-2610 (1987).

\bibitem{Weinberg1998}
H.\ Martel, P.\ R.\ Shapiro, S.\ Weinberg,
Likely values of the cosmological constant,
{\it The Astrophysical Journal}
{\bf 492}, 29-40 (1998).

\bibitem{Piran2016}
T.\ Piran, et al.,
Cosmic explosions, life in the universe, and the cosmological constant,
{\it Physical Review Letters\/} {\bf 116}, 081301 (2016).

\bibitem{Hoyle1954}
F.\ Hoyle, 
On nuclear reactions occurring in very hot stars. I.\ The synthesis of
elements from carbon to nickel,
{\it The Astrophysical Journal Supplement Series}
{\bf 1}, 121-146 (1954).

\bibitem{Carter1974}
B.\ Carter,
Large number coincidences and the anthropic principle in cosmology,
in {\it Confrontation of Cosmological Theories with Observational Data},
(Springer, Dordrecht, 1974) p.~291-298.

\bibitem{Adams2019}
Fred C.\ Adams,
The degree of fine-tuning in our universe -- and others,
{\it Physics Reports\/}
{\bf 807}, 1-111 (2019).

\bibitem{Dicke1961}
R.\ H.\ Dicke,
Dirac’s cosmology and Mach’s principle,
{\it Nature\/} {\bf 192}, 440-441 (1961).

\bibitem{Deutsch2011}
D.\ Deutsch, 
{\it The Beginning of Infinity: Explanations that Transform the World}
(Penguin, London, 2011), pp.~99-100.

\bibitem{Wang2023}
Z.-W. Wang,and S. L. Braunstein,
Sciama’s argument on life in a random universe and distinguishing apples from oranges,
{\it Nature Astronomy}, {\bf 7}, 1--2 (2023).

\bibitem{Milman1988}
V.\ D.\ Milman, 
The heritage of P.\ L\'evy in geometrical functional analysis,
{\it Ast\'erisque} {\bf 157}, 273-301 (1988).

\bibitem{Eddington1936}
A.\ S.\ Eddington,
{\it Relativity Theory of Protons and Electrons}
(Cambridge Univ.\ Press, Cambridge, 1936) p.~3, 

\bibitem{Siegel2018}
E.\ Siegel,
How many fundamental constants does it take to explain the universe?
Forbes, Nov.\ 16, 2018.

\bibitem{Eddington1939}
A.\ S.\ Eddington,
{\it The Philosophy of Physical Science}
(Cambridge Univ.\ Press, Cambridge, 1939).

\bibitem{Diaconis1984}
P.\ Diaconis \& D.\ Freedman, 
Asymptotics of graphical projection pursuit, 
{\it The Annals of Statistics}
{\bf 12}, 793-815 (1984).

\bibitem{Klartag2007}
B.\ Klartag, 
A central limit theorem for convex sets, 
{\it Inventiones Mathematicae} 
{\bf 168},  91-131 (2007).



\bibitem{Cirkovic2016}
M.\ M.\ \'Cirkovi\'c,
Fermi’s Paradox Is a Daunting Problem -- Under Whatever Label,
{\it Astrobiology}, {\bf 16}, 737-740 (2016).

\bibitem{SETI}
D. Overbye,
The Search for E.T.\ Goes on Hold, for Now,
{\it The New York Times}, March 23, 2020.

\bibitem{Wilde1889}
O.\ Wilde,
The decay of lying: A dialogue,
in {\it The Nineteenth Century: A Monthly Review},
ed.\ J.\ Knowles. Vol. XXV. January-June, 1889. pp.~35-56. 

\bibitem{Knaeble2015Ap}
B. Knaeble, 
Variations on the projective central limit theorem, 
{\it General Mathematics Notes\/}
{\bf 26(2)},  119-133 (2015).

\end{thebibliography}
\end{document}